# An Integral Measure of Aging/Rejuvenation for Repairable and Non-repairable Systems

M.P. Kaminskiy and V.V. Krivtsov

*Abstract* – This paper introduces a simple index that helps to assess the degree of aging or rejuvenation of a (non)repairable system. The index ranges from -1 to 1 and is negative for the class of decreasing failure rate distributions (or deteriorating point processes) and is positive for the increasing failure rate distributions (or improving point processes). The introduced index is distribution free.

## ACRONYMS[1]

| | |
|---|---|
| CDF | cumulative distribution function |
| CFR | constant failure rate |
| CIF | cumulative intensity function |
| DFR | decreasing failure rate |
| GPR | G-renewal process |
| HPP | homogeneous Poisson process |
| IFR | increasing failure rate |
| NHPP | non-homogeneous Poisson process |
| PP | point process |
| ROCOF | rate of occurrence of failures |
| RP | renewal process |
| TTF | time to failure |

## I. INTRODUCTION

In reliability and risk analysis, the terms *aging* and *rejuvenation* are used for describing reliability behavior of repairable as well as non-repairable systems (components). The repairable systems reliability is modeled by various point processes (PP), such as the homogeneous Poisson process (HPP), non-homogeneous Poisson process (NHPP), renewal process (RP), G-renewal process (GRP), to name a few. Among these PP, some special classes are introduced in order to model the so-called *improving* and *deteriorating* systems. An improving (deteriorating) system is defined as the system having decreasing (increasing) *rate of occurrence of failures* (ROCOF). It might be said that among the point processes used as models for repairable systems, the HPP (having a constant ROCOF) is a basic one, as the one modeling non-aging system reliability behavior.

Similarly, among the distributions used as models of time to failure (TTF) of non-repairable systems (components), the exponential distribution, which is the only distribution having a constant failure rate, plays a fundamental role. This distribution might be considered as the limiting between the class of *aging* or *increasing failure rate* (IFR) distributions and the class of *decreasing failure rate* (DFR) distributions. The distribution is closely related to the above mentioned HPP. Indeed, in the framework of the HPP model, the distribution of the intervals between successive events observed during a time interval [0, $t$] is the exponential one with parameter $\lambda$ equal to parameter $\lambda$ of the respective Poisson distribution with mean $\lambda t$.

In many practical situations, it is important to make an assessment how far a given point process deviates from the HPP, which can be considered as a simple and, therefore, strong competing model. Note that if the HPP turns out to be an adequate model, the respective system is considered as non-aging, so that it does not need any preventive maintenance (as opposed to the case, when a repairable system reveals aging).

The statistical tools helping to find out if the HPP is an appropriate model are mainly limited to statistical hypothesis testing, in which the null hypothesis is

$H_0$: "The times between successive events (*interarrival times*) are independent and identically exponentially distributed (i.e., the system is non-aging)", and the alternative hypothesis is

$H_1$: "The system is either aging or improving."

The most popular hypothesis testing procedures for the considered type of problems are the *Laplace test* [9] and the so-called *Military Handbook test* [7]. It should be noted that these procedures do not provide a simple measure quantitatively indicating how different the ROCOF of a given point process is, compared to the respective constant ROCOF of the competing HPP model.

Analogously, for the non-repairable units, there are some hypothesis testing procedures that help to determine if the exponential distribution is an appropriate TTF model. In such situations, in principle, any goodness-of-fit test procedure can be applied. Some of these tests for the null-hypothesis: "The times to failure are independent and identically exponentially distributed" appear to have good power against the IFR or DFR alternatives [6].

Among these goodness-of-fit tests, one can mention the G-test, which is based on the so-called *Gini statistic* [1]. In turn, the Gini statistics originates from the so-called *Gini coefficient* used in macroeconomics for comparing an income distribution of a given country with the uniform distribution covering the same income interval. The Gini coefficient is used as a measure of income inequality [10]. The coefficient takes on the values between 0 and 1. The closer the coefficient value to zero, the closer the distribution of interest is to the uniform one. The interested reader could find the index values sorted by countries in [5], that includes the UN and CIA data.

In the following sections, we introduce a Gini-type coefficient showing how fast a given non-repairable system is aging (rejuvenating) compared to the respective exponential distribution, having a "zero aging rate". The introduced coefficient takes on the values between -1 and 1. The closer the coefficient value to zero the closer the distribution of interest is to the exponential one. A positive (negative) value of the coefficient indicates an IFR (DFR) failure time distribution. Then, we introduce a similar coefficient for the repairable systems. This coefficient also takes on the values between -1 and 1. As in the previous case, the closer the coefficient value to zero, the closer the PP of interest is to the HPP. Analogously, a positive (negative) value of this coefficient will indicate that a given repairable system is deteriorating (improving). It should be noted that the suggested coefficient is only to a small extent similar to the Gini coefficient. For the sake of simplicity, in the following this Gini-type coefficient will be referred to as *GT coefficient* and denoted as *C*.

## II. GT COEFFICIENT FOR NON-REPAIRABLE SYSTEMS (COMPONENTS)

Consider a non-repairable system (component) whose TTF distribution belongs to the class of the IFR distributions. Denote the *failure rate* or the *hazard function* associated with this distribution by $h(t)$. The respective *cumulative hazard function* is then

$$H(t) = \int_0^t h(\tau)d\tau \qquad (1)$$

and is concave upward - see Figure 1.

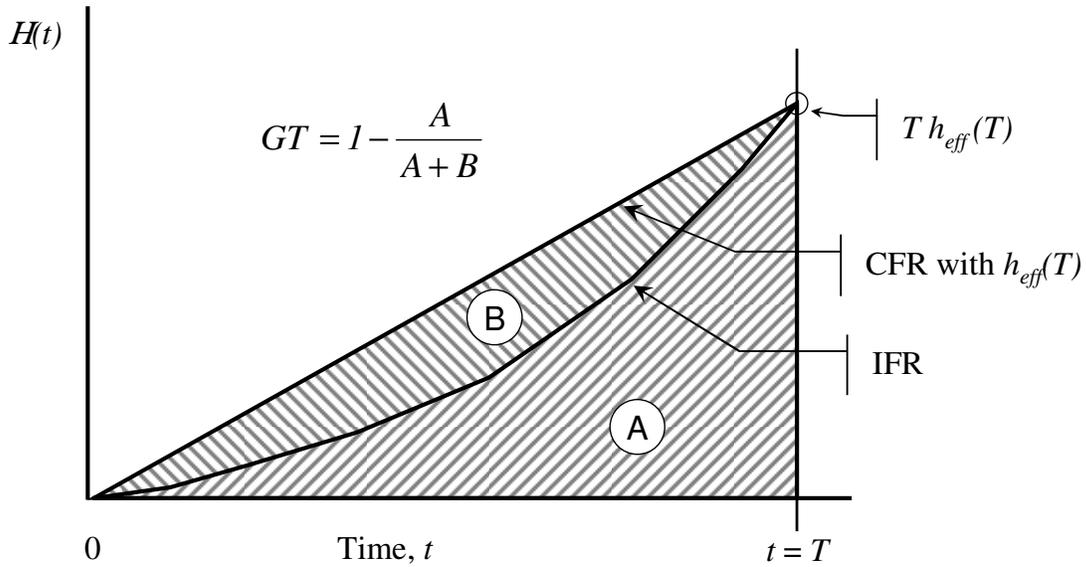

*Figure 1. Graphical interpretation of the GT coefficient for an IFR distribution.*

Consider time interval [0, T]. The cumulative hazard function at $T$ is $H(T)$, the respective CDF is $F(T)$ and the reliability function is $R(T)$. Now, introduce $h_{eff}$, as the failure rate of the exponential distribution with the CDF equal to the CDF of interest at the time $t = T$, i.e.,

$$h_{eff}(T) = -\frac{ln(1 - F(T))}{T} \qquad (2)$$

In other words, the introduced exponential distribution with parameter $h_{eff}$, at $t=T$, has the cumulative hazard function equal to the cumulative hazard function of the IFR distribution of interest, as depicted in Figure 1.

The GT coefficient, $C(T)$, is then introduced as

$$C(T) = 1 - \frac{\int_0^T H(t)dt}{0.5Th_{eff}(T)T} = 1 - \frac{2\int_0^T H(t)dt}{TH(T)} = 1 - \frac{2\int_0^T \ln(R(t))dt}{T\ln(R(t))} \qquad (3)$$

In terms of Figure 1, $C(T)$, is defined as one minus the ratio of areas $A$ and $A + B$. It is easy to check that the above expression also holds for the decreasing failure rate (DFR) distributions, for which $H(t)$ is concave downward.

It is clear that $C(T)$ satisfies the following inequality: $-1 < C(T) < 1$. The coefficient is positive for the IFR distributions, negative – for the DFR distributions and is equal to zero for the constant failure rate (CFR), i.e., exponential distribution. One can also show that the absolute value of $C(T)$ is proportional to the mean distance between the $H(t)$ curve and the $h_{eff} t$ line – see Figure 1. Note that the suggested coefficient is distribution-free.

*A. GT Coefficient for the Weibull Distribution*

For some TTF distributions, the GT coefficient can be expressed in a closed form. For example, in the most important (in the reliability context) case of the Weibull distribution with the scale parameter $\alpha$ and the shape parameter $\beta$, and the CDF of the form:

$$F(t) = 1 - \exp\left(-\left(\frac{t}{\alpha}\right)^\beta\right), \qquad (4)$$

the GT coefficient can be found as

$$C = 1 - \frac{2}{\beta + 1} \qquad (5)$$

It's worth noting that in this case, GT depends neither on the scale parameter $\alpha$, nor on time interval $T$. Also note that

$$C(\beta) = -C\left(\frac{1}{\beta}\right), \qquad (6)$$

which is illustrated by Table 1 below.

*Table 1. GT coefficient C for Weibull Distribution as Function of Shape Parameter β.*

| Shape Parameter β | C | TTF Distribution |
|---|---|---|
| 5 | 0.6(6) | IFR |
| 4 | 0.6 | IFR |
| 3 | 0.5 | IFR |
| 2 | 0.3(3) | IFR |
| 1 | 0 | CFR |
| 0.5 | -0.3(3) | DFR |
| 0.3 | -0.5 | DFR |
| 0.25 | -0.6 | DFR |
| 0.2 | -0.6(6) | DFR |

*B. GT Coefficient for the Gamma Distribution*

Although not as popular as the Weibull distribution, the gamma distribution still has many important reliability applications. For example, it is used to model a standby system consisting of $k$ identical components with exponentially distributed times to failure; the gamma distribution is also the conjugate prior distribution in Bayesian estimation of the exponential distribution.

Let's consider the gamma distribution with the CDF given by

$$F(t) = \frac{1}{\Gamma(k)} \int_0^{\lambda t} \tau^{k-1} e^{-\tau} d\tau = I(k, \lambda t), \qquad (7)$$

where $k > 0$ is the shape parameter, $1/\lambda > 0$ is the scale parameter, and $I(k,x) = \int_0^x y^{k-1} e^{-y} dy$ is the incomplete gamma function. Similar to the Weibull distribution, the gamma distribution has the IFR, if the shape parameter $k > 1$; DFR, if $k < 1$, and CFR, if $k = 1$.

Using definition (3), the GT coefficient for the gamma distribution can be written as

$$C(T) = 1 - \frac{2\int_0^T \ln(1 - I(k, \lambda \tau)) d\tau}{T \ln(1 - I(k, \lambda T))} \qquad (8)$$

Table 2 displays $C(T)$ for the gamma distribution with $\lambda = 1$ evaluated at $T = 1$.

*Table 2. GT Coefficient, C (T), for Gamma Distribution with λ =1 and T = 1.*

| Shape Parameter $k$ | $C(T)$ | TTF Distribution |
|---|---|---|
| 5 | 0.623 | IFR |
| 4 | 0.543 | IFR |
| 3 | 0.428 | IFR |
| 2 | 0.258 | IFR |
| 1 | 0.000 | CFR |
| 0.5 | -0.196 | DFR |
| 0.3 | -0.285 | DFR |
| 0.25 | -0.338 | DFR |
| 0.2 | -0.375 | DFR |

## III. GT COEFFICIENT FOR REPAIRABLE SYSTEMS

*A. Basic Definitions*

A point process (PP) can be informally defined as a mathematical model for highly localized events distributed randomly in time. The major random variable of interest related to such processes is the number of events, $N(t)$, observed in time interval [0, $t$]. Using the nondecreasing integer-valued function $N(t)$, the point process $\{N(t), t \geq 0\}$ is introduced as the process satisfying the following conditions:

1. $N(t) \geq 0$
2. $N(0) = 0$
3. If $t_2 > t_1$, then $N(t_2) \geq N(t_1)$
4. If $t_2 > t_1$, then $[N(t_2) - N(t_1)]$ is the number of events occurred in the interval $(t_1, t_2]$

The mean value $E[N(t)]$ of the number of events $N(t)$ observed in time interval *[0, t]* is called *cumulative intensity function* (CIF), *mean cumulative function* (MCF), or *renewal function*. In the following, the term *cumulative intensity function* is used. The CIF is usually denoted by $\Lambda(t)$:

$$\Lambda(t) = E[N(t)] \tag{10}$$

Another important characteristic of point processes is the *rate of occurrence of events*. In reliability context, the *events* are *failures,* and the respective rate of occurrence is abbreviated to ROCOF. The ROCOF is defined as the derivative of CIF with respect to time, i.e.

$$\lambda(t) = \frac{d\Lambda(t)}{dt} \quad (11)$$

When an event is defined as a failure, the system modeled by a point process with an increasing ROCOF is called *aging* (*sad*, *unhappy*, or *deteriorating*) system. Analogously, the system modeled by a point process with a decreasing ROCOF is called *improving* (*happy*, or *rejuvenating*) system.

The distribution of time to the first event (failure) of a point process is called the *underlying distribution*. For some point processes, this distribution coincides with the distribution of time between successive events; for others it does not.

*B. GT Coefficient*

The suggested below measure of non-homogeneity of occurrence of events for the sake of simplicity and consistency with Section II is further referred to as GT coefficient, and denoted by *C*. The coefficient is introduced as follows.

A PP having an integrable over *[0, T]* cumulative intensity function, $\Lambda(t)$, is considered. It is assumed that the respective ROCOF exists, and it is increasing function over the same interval *[0, T]*, so that $\Lambda(t)$ is concave upward as illustrated by Figure 2. Introduce the HPP with CIF $\Lambda_{HPP}(t) = \lambda t$ that coincides with $\Lambda(t)$ at *t = T*, i.e., $\Lambda_{HPP}(T) = \Lambda(T)$, – see Figure 2. For the given time interval *[0, T]* the GT coefficient is defined as

$$C(T) = 1 - \frac{2\int_0^T \Lambda(t)\,dt}{T\Lambda(T)} \qquad (12)$$

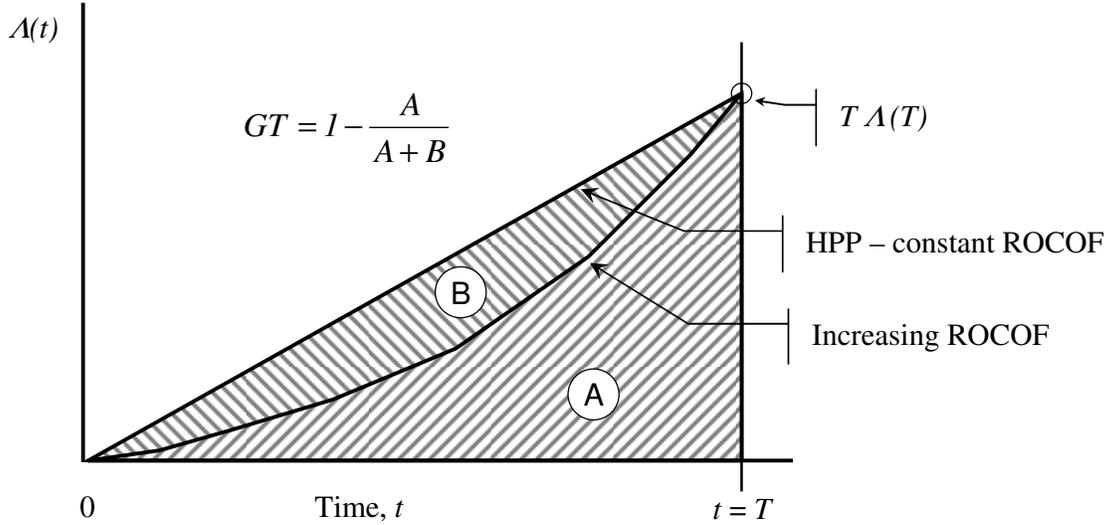

*Figure 2. Graphical interpretation of GT coefficient for an increasing ROCOF point process.*

It is obvious that for a PP with an increasing ROCOF, the GT coefficient is positive and for a PP with a decreasing ROCOF, the coefficient is negative. The smaller the absolute value of the GT coefficient, the closer the considered PP is to the HPP.

Clearly, for the HPP, $C(T)=0$. GT coefficient satisfies the following inequality: $-1 < C(T) < 1$. One can also show that the absolute value of GT coefficient $C(T)$ is proportional to the mean distance between the $\Lambda(t)$ curve and the CIF of the HPP.

For the most popular NHPP model – the *power law* model with the underlying Weibull CDF (4) – the GT coefficient is expressed in a closed form:

$$C = 1 - \frac{2}{\beta + 1} \qquad (13)$$

Note that (13) is exactly the same as (5). This is because NHPP's CIF is formally equal to the cumulative hazard function of the underlying failure time distribution (see, e.g., [4]).

Some examples of applying the GT coefficient to various PP commonly used in reliability and risk analysis are given in Table 3. *Repair effectiveness factor* in Table 3 refers to the degree of restoration upon the failure of a repairable system; see [3], [2]. This factor equals zero for an RP, one – for an NHPP and is greater-or-equal-to zero – for a GRP (of which the RP and the NHPP are the particular cases).

*Table 3. GT coefficients of some PP over time interval [0, 2].*
*Weibull with scale parameter α=1 is used as the underlying distribution.*

| Stochastic Point Process | Shape parameter of Underlying Weibull Distribution | Repair Effectiveness Factor | GT Coefficient, $C$ |
|---|---|---|---|
| HPP | 1 | N/A | 0 |
| NHPP | 1.1 | 1 | 0.05 |
| NHPP | 2 | 1 | 0.33 |
| NHPP | 3 | 1 | 0.50 |
| RP | 2 | 0 | 0.82 |
| GRP | 2 | 0.5 | 0.21 |

Note: the GT coefficient for RP and GRP was obtained using numerical techniques.

*About the authors*

**Mark Kaminskiy** is the Chief Statistician at the Center of Technology and Systems Management of the University of Maryland (College Park), USA. Dr. Kaminskiy is a researcher and consultant in reliability engineering, life data analysis and risk analysis of engineering systems. He has conducted numerous research and consulting projects funded by the government and industrial companies such as Department of Transportation, Coast Guards, Army Corps of Engineers, US Navy, Nuclear Regulatory Commission, American Society of Mechanical Engineers, Ford Motor Company, Qualcomm Inc, and several other engineering companies. He taught several graduate courses on Reliability Engineering at the University of Maryland. Dr. Kaminskiy is the author and co-author of over 50 publications in journals, conference proceedings, and reports.

**Vasiliy Krivtsov** is a Senior Staff Technical Specialist in reliability and statistical analysis with Ford Motor Co. He holds M.S. and Ph.D. degrees in Electrical Engineering from Kharkov Polytechnic Institute, Ukraine and a Ph.D. in Reliability Engineering from the University of Maryland, USA. Dr. Krivtsov is the author and co-author of over 50 professional publications, including a book on *Reliability Engineering and Risk Analysis*, 9 patented inventions and 2 Ford corporate secret inventions. He is an editor of the Elsveir's *Reliability Engineering and System Safety* journal and is a member of the IEEE Reliability Society. Prior to Ford, Krivtsov held the position of Associate Professor of Electrical Engineering in Ukraine, and that of a Research Affiliate at the University of Maryland Center for Reliability Engineering. Further information on Dr. Krivtsov's professional activity is available at *www.krivtsov.net*